\documentclass[12]{article} 

\usepackage{amsfonts}
\usepackage{graphicx}
\usepackage{dcolumn}
\usepackage{bm}
\usepackage{amsmath}
\usepackage[version=3]{mhchem}

\newcommand{\vett}[1]{\mathbf{#1}} 
\newcommand{\reali}{\mathbb{R}}
\newcommand{\eps}{\varepsilon}
\newcommand{\acknowledgments}{\vskip1.em\noindent\textbf{Acknowledgments}.\ }
\newcommand{\Name}[1]{\textsc{#1}, }
\newcommand{\Book}[1]{\textsl{#1}}
\newcommand{\Publ}[1]{{\normalfont(#1)}}
\newcommand{\Year}[1]{(#1)}
\newcommand{\Page}[1]{\unskip,\ {\normalfont p.~#1}}  
\newcommand{\REVIEW}[4]{\textit{#1} \textbf{#2}, (#3) #4}

\title{Infrared optical properties of $\alpha$ quartz by molecular
  dynamics simulations  }

\author{Fabrizio Gangemi\thanks{
          DMMT, Universit\`a di Brescia, Viale Europa 11, 
	  25123 Brescia -- Italy} \and 
        Roberto Gangemi\footnotemark[1]  \and
        Andrea Carati\thanks{
	  Dep. Mathematics, Universit\`a degli Studi di  Milano, 
	  Via Saldini 50, 20133 Milano -- Italy } \and
        Alberto Maiocchi\footnotemark[2] \and
        Luigi Galgani\footnotemark[2] 
}

\begin{document}

\maketitle

\begin{abstract}
  This paper is concerned with  theoretical estimates of
  the refractive--index curves for quartz, obtained by 
  the Kubo {formul\ae} in  the classical approximation, through
  MD simulations for the motions of the ions. Two
  objectives are considered. The first one is to understand the role
  of nonlinearities in situations where they are very large, as at the
  $\alpha$--$\beta$ structural phase transition. We show that on the
  one hand they don't play an essential role in connection with the
  form of the spectra in the infrared. On the other hand they play an
  essential role in introducing a chaoticity which involves a definite
  normal mode. This might explain why that mode is Raman active in the
  $\alpha$ phase, but not in the $\beta$ phase. The second objective
  concerns whether it is possible in a microscopic model to obtain
  normal mode frequencies, or peak frequencies in the optical spectra,
  that are in good agreement with the experimental data for
  quartz. Notwithstanding a lot of effort, we were unable to find
  results agreeing better than about 6\%, as apparently also occurs
  in the whole available literature. We interpret this fact as
  indicating that some essential qualitative feature is lacking in all
  models  which consider, as the present one, only short--range 
  repulsive potentials and unretarded long--range electric forces.  
\end{abstract}
\textbf{PACS}: 78.20.Ci, 42.70.Ce, 63.20.Ry

\section{Introduction}

A subject of great current interest is that of a microscopic description of
the ferroelectric transition. It is known that at the transition a
divergence of the dielectric constant $\eps(\omega)$ at $\omega=0$
occurs, which  in most cases is understood as corresponding to the
fact that the frequency of an infrared peak goes to zero.
 On the other hand, the infrared frequencies are usually
studied via a linear analysis of phonon dispersion relations, while
the nonlinear contribution to the dynamics should be relevant at the
transition to the ferroelectric phase, as should be near any phase
transition. So the problem arises 
of how the infrared peaks should be described in a fully nonlinear
setting. A description can actually be given  through the study of the time
autocorrelation of the polarization due to the ions, which can be
computed in the classical approximation via molecular dynamics
simulations. Computations of this type were indeed performed
successfully in the case of \ce{LiF} \cite{noi}, with results that agree
with the experimental data in a surprisingly good way.

At the moment, as will be explained below, we are unable to study
ferroelectrics through molecular dynamics. So in this paper we limit
ourselves to quartz, which is not a ferroelectric material, but
however presents a divergence in 
the dielectric constant at the temperature of the $\alpha$--$\beta$
transition. We compute here  its refractive index curves in the
infrared region.  Our main concerns are the dynamical properties of
the system, particularly at the $\alpha$--$\beta$ transition, and the
quantitative agreement between calculated and 
experimental spectra in the infrared for $\alpha$ quartz.

As is well known \cite{spitzer}, linear analysis shows that quartz is
doubly refractive and that the peaks in its refractive--index curves
correspond to the frequencies of the active normal modes. Such
qualitative results are confirmed by the present molecular dynamics
simulations for $\alpha$ quartz at high temperatures, even at the
transition to the $\beta$ phase, notwithstanding the high nonlinearity
of the system.  This quantitative agreement of the nonlinear results
with those of the linear analysis, in particular for the values of the
frequencies, is quite surprising, in view of the large nonlinear
contributions, and probably is due to some deep reason not yet fully
understood.  On the other hand, both the linear analysis and our
nonlinear study fail in perfectly reproducing the experimental data,
as some systematic deviations are observed. We
tried several procedures for choosing the parameters, both of the
linear model and of the nonlinear one, in order to find a better
agreement with the experimental curves. But there was no way of
reducing the relative error below a threshold of the order of
6\%. This fact, too, requires an explanation.  These are the two main
results of the present work.

\section{The model}
\begin{figure}
  \begin{center}
    \includegraphics[width = 0.9\textwidth]{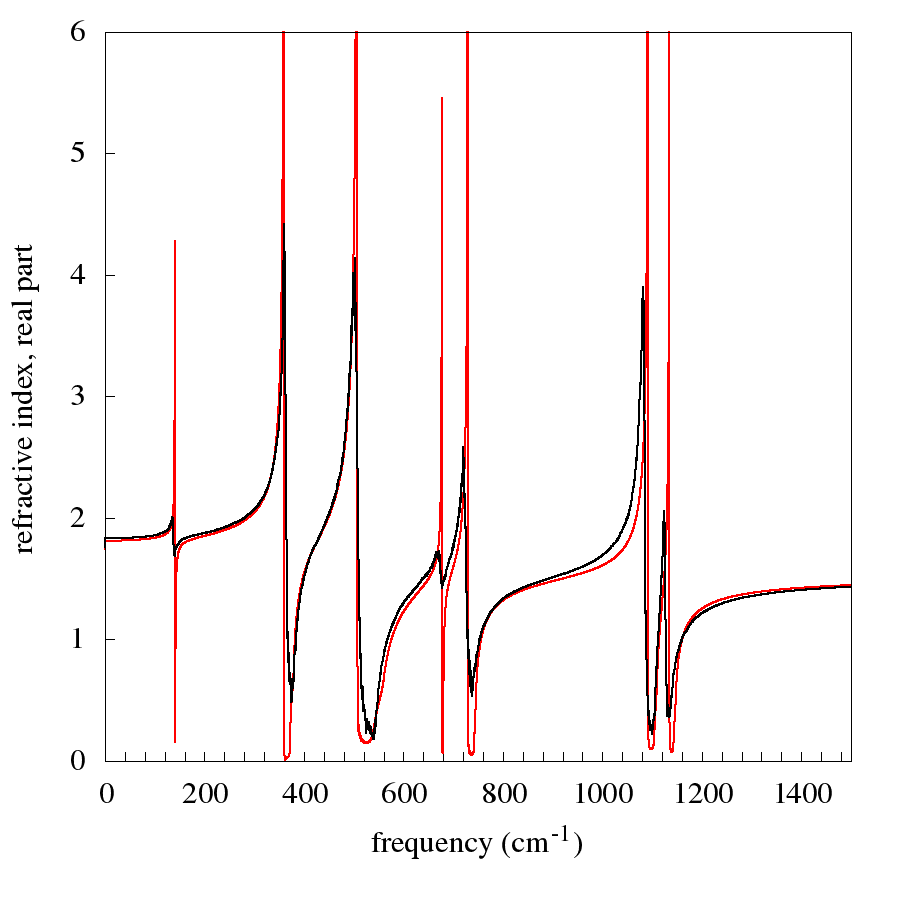}
  \end{center}
  \caption{\label{fig:1} Refractive index (real part) versus  $\omega$ for the
    ordinary ray, at temperatures $T= 1\,$K (red line) and  
    $T= 300\,$K (dark line). Color online.}
\end{figure}
\begin{figure}
  \begin{center}
    \includegraphics[width = 0.9\textwidth]{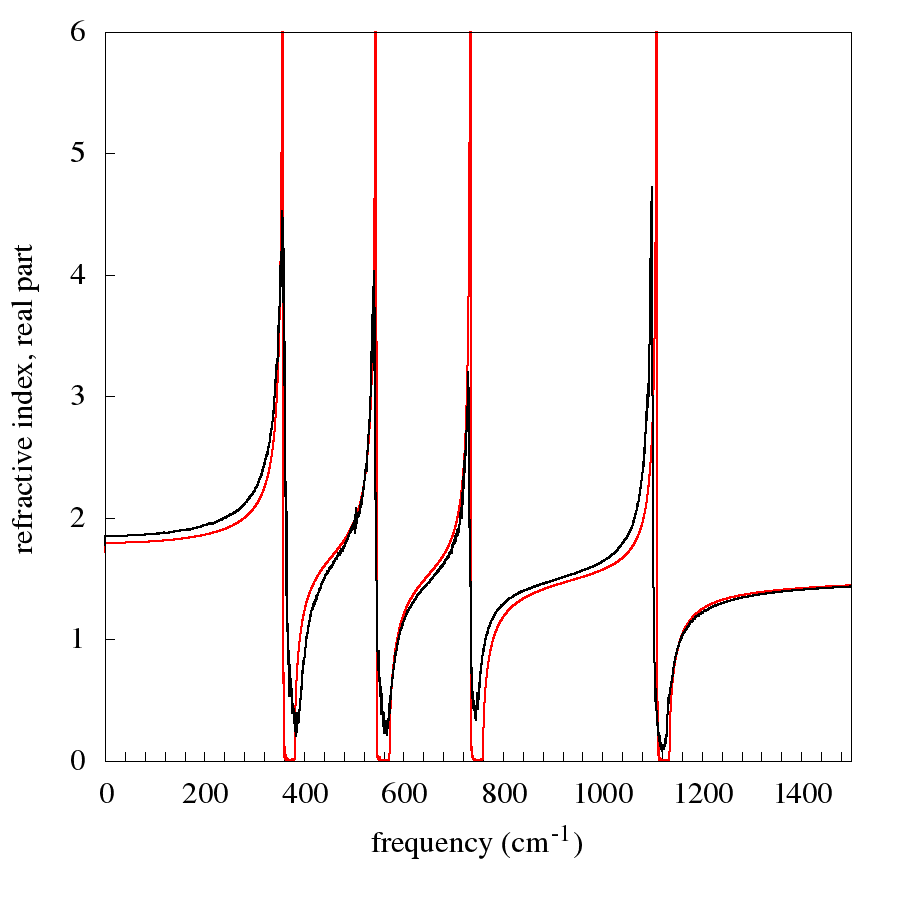}
  \end{center}
  \caption{\label{fig:2} Same as Figure \ref{fig:1} for the
    extraordinary ray.}
\end{figure}
\begin{table}[bh]
  \begin{center}
    \begin{tabular}{cccc}
      \hline
      && $\alpha$ quartz parameters & \\
      \hline
      & $a$ & $b$ & $c$ \\
      & 4.9137 \AA\ & 4.9137 \AA\ & 5.4047 \AA\ \\
      \hline
      & $x$ & $y$ & $z$ \\
      \ce{Si}\ & 0.4697\ & 0.0000\ & 0.0000\ \\
      \ce{O} & 0.4133 & 0.2672 & 0.1188 \\ 
      \hline
      && $\beta$ quartz parameters &\\
      \hline
      & $a$ & $b$ & $c$ \\
      & 4.9965 \AA & 4.9965 \AA & 5.4546 \AA \\
      \hline
    & $x$ & $y$ & $z$\\
      \ce{Si}\ & 0.5000\ & 0.0000\ & 0.0000\\
      \ce{O} & 0.4157 & 0.2078 & 0.1667 \\
      \hline
    \end{tabular}
  \end{center}
  \caption{\label{tab:1} Geometric parameters for $\alpha$ and
    $\beta$ quartz (see text).}
\end{table} 

It is known that the primitive cell of quartz contains nine
atoms (three silicon and six oxygen atoms). Moreover, it has the form
of a right prism of rhomboidal basis, corresponding to three basis
vectors ${\bf a}, {\bf b}, {\bf c}$ with ${\bf a}$ and ${\bf b}$
forming an angle of 120$^\circ$ and ${\bf c}$ orthogonal to ${\bf a}$
and ${\bf b}$. In Table~\ref{tab:1} we report the lengths of
the three basis 
vectors at  normal conditions of temperature and pressure ($300\,
\mathrm{K}$ and 1 bar), as given by \cite{kihara}, which we use in our
simulations. We also report the fractional coordinates $x$,$y$,$z$
(along the three basis vectors) of a silicon
atom and of an oxygen atom, out of which all other coordinates can be
generated by symmetry transformations.\footnote{We recall 
(see \cite{spitzer}) that the space   group of $\alpha$ quartz is
$P3_121$ or $P3_221$. Its transformations are the result of a rotation
belonging to the dihedral point group $D_3$ and a translation of a
multiple of $2/3\cdot{\bf c}$. The group has three irreducible 
representations, usually denoted as $A_1$
   (totally symmetric, one-dimensional), $A_2$ (one-dimensional) and
$E$ (two-dimensional).} The configuration corresponding
to $\alpha$ quartz is thought of as being the more stable equilibrium
configuration of the system.
In order to simulate the crystal, we choose
a domain $D\subset \reali^3$ (fundamental box) constituted by $4\times
4\times 4$ primitive cells, with a number $N=9\times 4^3=576$ of point
particles inside it. Due to the partially ionic character of the
quartz crystal, the point particles  have to be endowed with
suitable effective charges, $e_S$ for the silicon ion and $e_O$ for
the oxygen ion, with the neutrality constraint
\begin{equation}\label{neutro}
  2e_O+e_S=0\ .
\end{equation}
Thus, Coulomb long range forces come into play and, in order to take
them into account, working however with a small number $N$ of
particles, periodic boundary conditions are imposed.  In addition to
the electric forces, short-range two-body spherically symmetric
potentials are introduced, one for each of the pairs \ce{Si-Si}, \ce{O-O},
\ce{Si-O}.  These potentials are taken of a form which is extensively used
for quartz, namely (see \cite{TTAM,BKS}),
\begin{equation}\label{potenziale}
  U(r)= Ae^{-Br}-\frac{ C}{r^{6} }\ ,
\end{equation}
($r$ being the interatomic distance), with a triple of parameters $A$,
$B$, $C$ a priori different for each pair.
In the numerical calculations, for the short-range interactions a
cutoff of 9 \AA\ was imposed, while the Coulomb interactions were
dealt with through standard Ewald summations. This is exactly the point
where some modifications are required if one aims at dealing with
ferroelectrics.  Indeed the Ewald resummation provides a periodic electric
field having a vanishing mean, at variance with
what occurs with ferroelectrics.  The problem of how to modify
accordingly the Ewald procedure is left for a possible future work.
\begin{figure}
  \begin{center}
    \includegraphics[width = 0.9\textwidth]{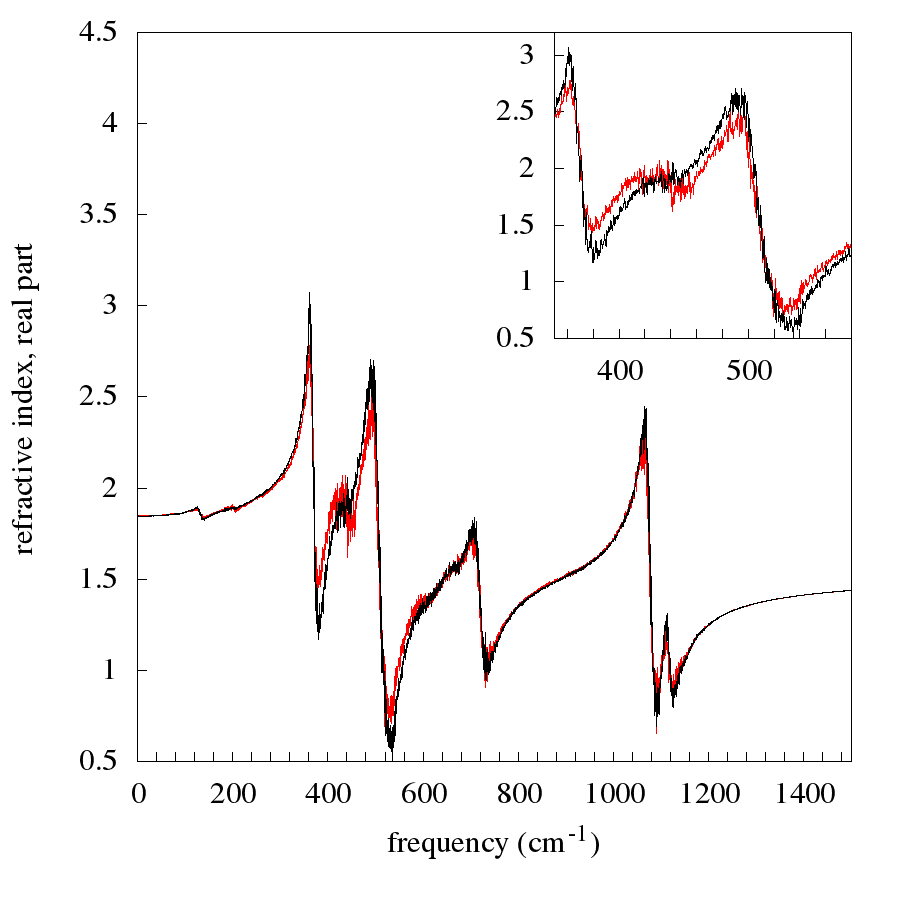}
  \end{center}
  \caption{\label{fig:4} Refractive index (real part) versus  $\omega$ for the
    ordinary ray, at temperature $T= 670\,$K, below
    the phase transition (red line), and at $T= 700\,$K,  above it 
    (dark line). The
    inset exhibits the new peak at about 455 cm$^{-1}$ in the $\beta$
    phase. Color online.}
\end{figure}
\begin{figure}
  \begin{center}
    \includegraphics[width = 0.9\textwidth]{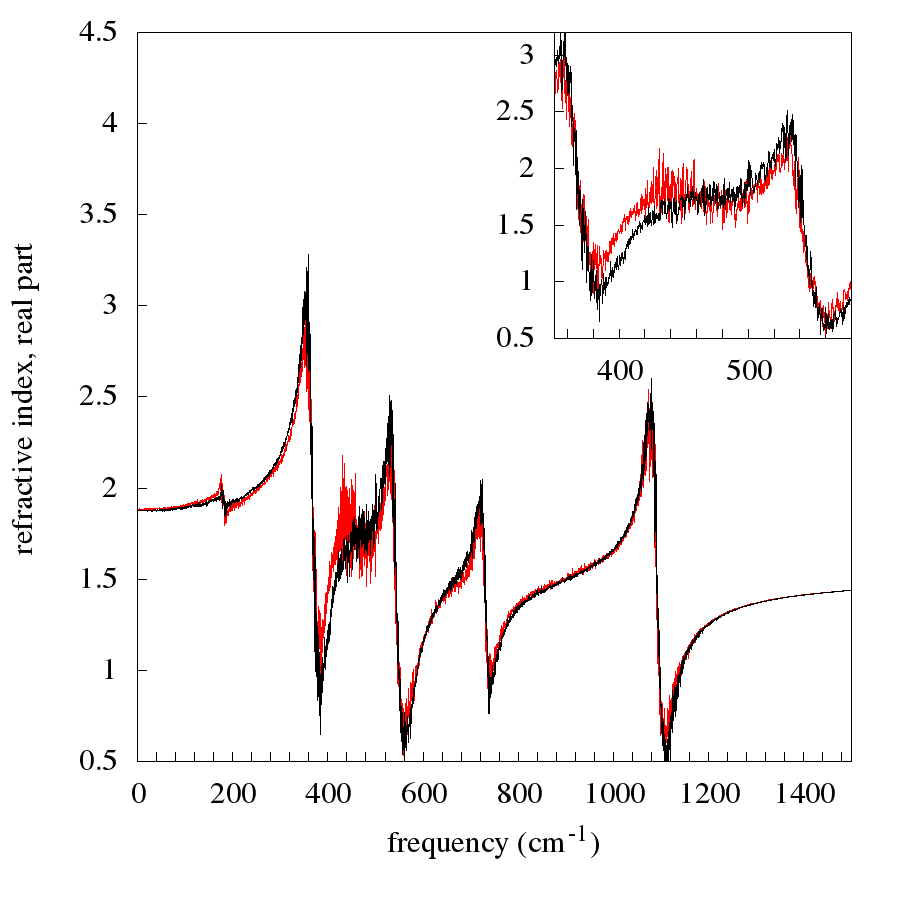}
  \end{center}
  \caption{\label{fig:5}  Same as Figure \ref{fig:4} for the
    extraordinary ray.}
\end{figure}

The masses are taken from the literature, so that to fix the model
there remains a total of 10 free parameters: one effective charge (for
example that of oxygen) and the three parameters $A$, $B$, $C$ 
of the short-range
potential for each of the three pairs \ce{Si-O}, \ce{O-O},
\ce{Si-Si}. The values of the potential parameters adopted
are reported in Table~\ref{tab:2}, while we used the values $e_S= 2.04191$
and  $e_O=-1.02096$ for the effective charge (in unit of electron
charge) of \ce{Si} and
\ce{O} respectively. They were determined by optimization
procedures aimed at obtaining the best possible agreement between the
computed refractive index curves  and the empirical ones at
300 K, requiring in addition that the $\alpha$ structure be stable at
that temperature and at larger (but not too much) temperatures.
With the values thus determined for the parameters of the model it
turns out,  as will be shown below, that the $\alpha$--$\beta$
transition occurs at a temperature of about $T= 700\,$K, which
is a rather low value. We leave for a future work the task of
performing  an optimization of  the parameters with a procedure  which 
takes both the frequencies   and the transition
temperature into account.
\begin{table}[bh]
  \begin{center}
    \begin{tabular}{cccc}
      & $A$ (eV) & $B$ (\AA$^{-1})$ & $C$ (eV \AA$^{6})$\\
      \hline
      \\[-10.pt]
      \ce{Si-O} &  18207.1 &  4.88538 &
      135.021  \\ 
     \ce{O-O} & 501.814 & 2.76745 & 15.0427\\
     \ce{Si-Si}  & 25.3672 & 1.41444 &  0.694560
    \end{tabular}
  \end{center}
  \caption{\label{tab:2} 
    Parameters of the short range potential. }
\end{table}

The equations of motion were numerically solved using a Verlet
algorithm, with integration step of 2 fs,  at
several values of temperature.
Having chosen to work in a purely Hamiltonian frame,
 temperature was  determined through the choice of the
initial data. In principle this should be obtained by extracting the
data according to a Gibbs distribution. This being impracticable, we
followed the alternative standard procedure. Namely, one puts the
particles at the $\alpha$ equilibrium point, 
while their velocities are
extracted from a Maxwell--Boltzmann distribution at a suitable
temperature. Then one lets the system thermalize, which usually takes
a time of the order of 2 ps (1000 integration steps), and temperature
is eventually 
identified through the mean kinetic energy of the ions. Then one
starts computing means and correlations of the relevant quantities.

\section{The refractive--index curves}
The refractive index is obtained by computing the electric
permittivity tensor $\eps_{ij}(\omega)$ as a function of frequency,
and diagonalizing it at each given frequency. As expected, two
eigenvalues are found to coincide, and the square root of such a value
is precisely the refractive index of the ordinary ray.  The refractive
index of the extraordinary ray is instead the square root of the
remaining eigenvalue.\footnote{It may be useful to keep to the
  following criterion: the eigenvalue corresponding to the eigenvector
  with larger component along the c-axis of the lattice is always
  associated with the extraordinary ray.}

The connection with dynamics is obtained through the susceptibility
tensor $\chi_{ij}(\omega)$ due to the ions, which is related to
permittivity by
\begin{equation}\label{perme}
\eps_{ij}(\omega)=\delta_{ij}+4\pi(\chi_{ij}\left(\omega)+
\chi^{el}_{ij}\right)\ .
\end{equation}
Here, $\chi^{el}$ is the contribution of the electrons, which turns
out to be  constant in the infrared domain
(see  \cite{palik}).  Instead,
 the ions' contribution $\chi_{ij}(\omega)$ is
obtained numerically according to Green-Kubo linear response theory
(see for example \cite{cg}) as follows.  One considers the
polarization $\vett P$, which is defined in microscopic terms as
\begin{equation}\label{pola}
\vett P= \frac 1{V} \sum_{l} e_l \vett x_{l}\ ,
\end{equation}
where $V$ is the volume of the simulation domain (or fundamental box),
while $\vett x_{l}$ is the position vector of the $l$--th ion, of
charge $e_l$.  
Then at temperature $T$ one has
\begin{equation}\label{kubo}
\chi_{ij}(\omega) = \frac V{k_BT}\int_0^{+\infty} e^{-i\omega t} \langle
P_i(t) \dot P_j(0)\rangle d t \ ,
\end{equation}
 $k_B$ being the Boltzmann constant. Here $\langle \cdot \rangle $
should in principle be the canonical average.  Actually the averages
were estimated as the mean of the time averages calculated along a
certain number (usually 40) of different MD trajectories, calculated
for 200 ps.

A first set of results is illustrated in Figures~\ref{fig:1} and
\ref{fig:2}. In the first figure we report, vs frequency,  the real
part of the 
refractive index for the ordinary ray at $T=300\, \mathrm{K}$
(dark line) and at $T=1\; \mathrm{K}$ (red line).
The analogous spectra for the extraordinary ray are
reported in Figure~\ref{fig:2}. At a
temperature as low as $T=1\; \mathrm{K}$ the spectrum is
determined essentially by the linear approximation, so that the peaks
correspond to the frequencies of the normal modes (actually, those of
the so called type $A_2$ and $E$; see \cite{spitzer}).  The results
show that the spectrum at $T=300\,\mathrm{K}$ does not differ
essentially from that corresponding to $1\; \mathrm{K}$, apart
from a consistent broadening of the peaks and some small shifts in their
frequencies.  

A second set of results concerns the behavior of the spectra at the
structural  $\alpha$--$\beta$ phase transition  which, in
the present microscopic model, with the choice made for the
parameters, turns out to occur in a region of temperatures roughly 
around $T=690\,$K. This will  be shown in a moment.
So we computed the refractive--index curves at 
$T=670\,$K, an $T=700\,$K, which are reported in Figure~\ref{fig:4}
for the ordinary ray  and  in Figure \ref{fig:5} for the
extraordinary ray.

The figures show that, at the transition, the optical spectra are
still dominated by the linear behavior.  Indeed, in both figures the
two curves relative to the two temperatures essentially superpose one
another, and one can notice the same peaks of the previous Figures
\ref{fig:1} e \ref{fig:2}, just a little more broadened and noisy.
\begin{figure}
  \begin{center}
    \includegraphics[width = 0.9\textwidth]{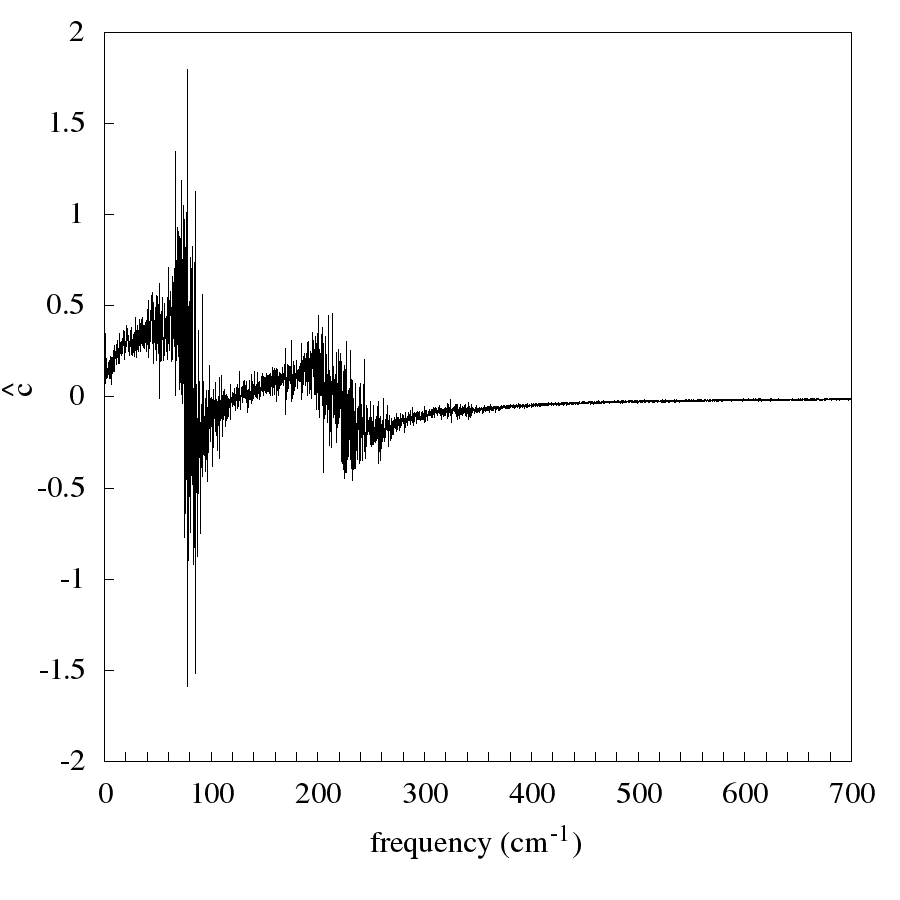}
  \end{center}
  \caption{\label{fig:6} Fourier transform $\hat C$ of the
    time--autocorrelation of the amplitude of the mode $\omega=232$
    cm$^{-1}$, at temperature $T=690\,\mathrm{K}$.
  }
\end{figure}

Some differences however show up. The most important one is the
appearance of 
one more peak (see the inset)  in the $\beta$ phase at
$\omega \simeq 455$ cm$^{-1}$, that should correspond to a normal mode which
is only Raman active in the $\alpha$ phase. In addition, a small peak
appears at approximately 200 cm$^{-1}$ in the extraordinary ray, which
should  presumably  be due to the nonlinearity. 

So the harmonic approximation essentially still dominates up to the
transition temperature, at least for what concerns the refractive
index.  On the other hand the transition has a relevant effect on at
least one of the normal modes of the system, the one corresponding to
the normal mode frequency $\omega=206$ cm$^{-1}$ (not active in the
infrared) which, with the choice made for the parameters of our model,
occurs at about $\omega=232$ cm$^{-1}$. Indeed such a mode exhibits a
chaotic behavior precisely at the transition. This can be seen from
Figure~\ref{fig:6}, where the Fourier transform  $\hat C$ of the
time--autocorrelation of the mode amplitude at temperature 690 K is
reported vs frequency.  One sees that the peak in question disappears,
being replaced by a very complex structure, characteristic of a
chaotic motion.  This fact might entail  an effect on the Raman
spectrum.  Indeed the mode in question is known to be
Raman active in the $\alpha$ quartz but not in the $\beta$ quartz, and
this seems to be explained by the chaoticity exhibited here at the
transition.  We leave for future work a detailed analysis of the
behavior of the normal modes at the transition.

\section{The $\alpha$--$\beta$ transition}
The occurrence of the transition is exhibited  in terms of an
``order parameter'', which discriminates between the configurations of the
two phases. Following essentially \cite{tsune} and \cite{Ma}, 
we define it as follows.
\begin{figure}
  \begin{center}
    \includegraphics[width = 0.9\textwidth]{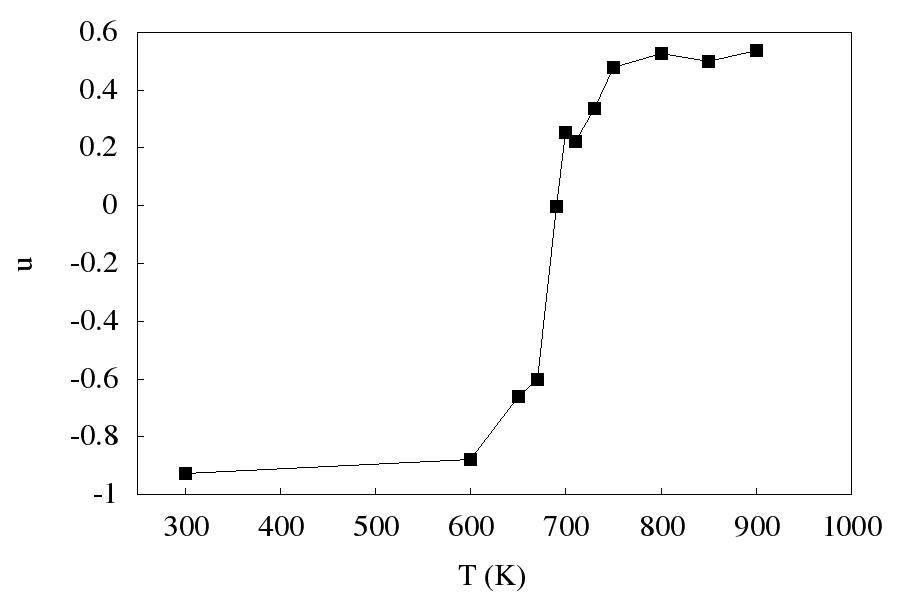}
  \end{center}
  \caption{\label{fig:3} The order parameter $u$ (see eq. 
    \ref{orderp}) as a function of $T$.
    A negative value corresponds to the $\alpha$ quartz phase, while 
    a positive value corresponds to the $\beta$ quartz phase. A rather
    abrupt change is exhibited near $T\simeq 690\,$K. }
\end{figure}

Consider the representative position vectors of silicon and of oxygen,
defined through their mean fractional coordinates (i.e., as the
averages, over the elementary cells, of the  fractional
coordinates of such atoms). Consider also their time averages
(actually a mean of such averages, taken over several independent
simulations), which we denote by $\vett x_S$, $\vett x_O$.  Then
consider the position vectors $\vett x^\alpha_S$ and $\vett
x^\alpha_O$, defined by the experimental fractional coordinates of
silicon and of oxygen for $\alpha$ quartz, taken from
Table~\ref{tab:1}.  Analogously define $\vett x^\beta_S$ and $\vett
x^\beta_O$.

Thus  the distance $d_{\alpha}$ of the mean configuration
of the system from  the equilibrium $\alpha$ configuration is
naturally estimated as
$$
d_{\alpha}=\sqrt{\|\vett x_S - \vett x^\alpha_S\|^2 + \|\vett x_O - \vett
  x^\alpha_O\|^2}\ , 
$$
and analogously for the distance $d_{\beta}$ from the equilibrium 
$\beta$ configuration. So one can introduce the variable $u$ defined by
\begin{equation}\label{orderp}
 u  =  \frac{d_{\alpha}-d_{\beta}}{d_{\alpha, \beta}}
\end{equation}
where ${d_{\alpha, \beta}}$ is a normalizing factor, the distance
between the two equilibria, defined in the natural way.

A negative value of $u$  clearly indicates that the atoms are, in
the mean, near to the $\alpha$ configuration, while a positive value
indicates that they are in the mean near to the $\beta$ configuration.
The graph of $u$ vs temperature, is reported in
Figure~\ref{fig:3}. One sees that a rather abrupt passage from a negative to
a positive value occurs in a small region of temperatures about
$T=690\, \mathrm{K}$, at which a value of $ u$ very near to
zero is obtained. Instead a value of about $-0.6$ is obtained at
$T=670\, \mathrm{K}$ and a value of about $+0.25$ is obtained at
$T=700\,\mathrm{K}$.  This shows that at these temperatures the
nonlinear effects become so important as to trigger a phase
transition.

\section{Comparison with the experimental data}
\begin{figure}[t]
  \begin{center}
    \includegraphics[width = 0.9\textwidth]{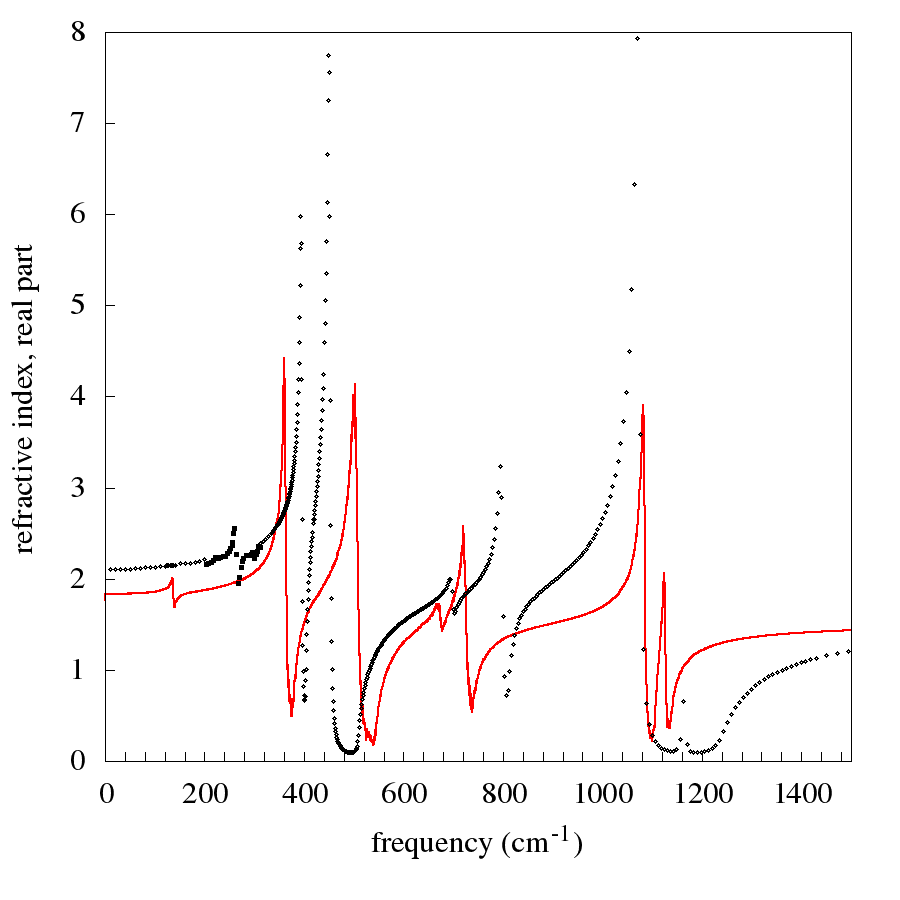}
  \end{center}
  \caption{\label{fig:7} 
  Real part of the refractive index  for the ordinary ray as a 
  function of $\omega$, at a temperature of 300 K.
  Comparison of the calculated curve (red line) with the experimental
  data, taken from ref.~\cite{palik} (diamonds) and ref.~\cite{cummings}
  (squares). Color online.}
\end{figure}

\begin{figure}[t]
  \begin{center}
    \includegraphics[width = 0.9\textwidth]{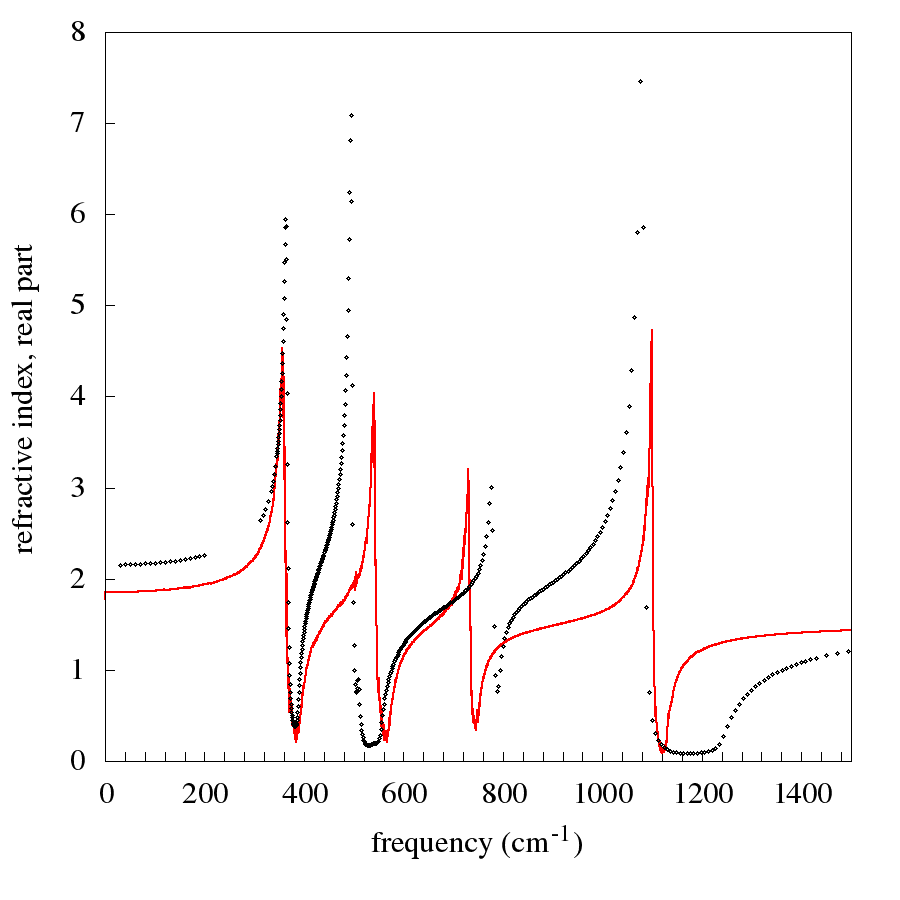}
  \end{center}
  \caption{\label{fig:8} Same as Figure \ref{fig:7} for the
    extraordinary ray. Here, however,  the experimental   data
(diamonds) are   taken only from ref.~\cite{palik}.}
\end{figure}

In Figures~\ref{fig:7} and \ref{fig:8} we report both  the calculated
refractive--index curves and the experimental ones,  taken 
from \cite{palik,cummings}, for the ordinary
and the extraordinary rays respectively, at $300\,
\mathrm{K}$. For both types of rays the experimental and the
calculated curves have the same general aspect: namely, the number of
peaks is the same, and both the intensities and the broadening are of
the same order. Actually the lowest peak in the theoretical curve
corresponds to the normal mode at 140 cm$^{-1}$, which has a vanishingly
small intensity in the data; on the contrary the lowest frequency peak
in the experimental curve is at 265 cm$^{-1}$ and the corresponding
mode in the theoretical curve has a very low intensity (see the normal
modes frequencies  in Table~\ref{tab:3}). 

However there is a clear quantitative
disagreement,  
determined essentially by the positions of the peaks. As in our model
we have ten free parameters, one can investigate whether a better
quantitative agreement can be obtained by optimizing them, or even by
considering other types of models.  Notwithstanding a lot of effort,
we were unable to significantly improve the agreement.

We now describe the strategy we followed for optimizing the
parameters.  The procedure is quite involved, because we have two
objectives.  On the one hand the system has to admit a global
equilibrium configuration, periodic with respect to the primitive cell
and furthermore reproducing the $\alpha$ quartz symmetries.  On the
other hand
we require that the normal mode frequencies, calculated 
at the $\alpha$ 
equilibrium, reproduce the frequencies observed, both in the infrared
spectra and in the Raman ones.

Our procedure was the following one. To start up, we consider the experimental
crystallographic configuration  of Table 
\ref{tab:1}, given by X ray diffraction, and  linearize the equations 
of motions at that point. So we can determine the 
 normal modes frequencies at the corresponding equilibrium point, 
as functions of the parameters entering the potential.
In this calculation, the
symmetries of the 
crystal are automatically taken into account in the construction of
the dynamical matrix, because only some components are directly
calculated, all the others being derived by symmetry
transformations. As a consequence, the modes are correctly grouped
into the three irreducible representations of the symmetry group,
namely 4 $A_1$ modes, 4 $A_2$ modes and 8 degenerate $E$ modes, so that
a total of 16 distinct frequencies are obtained. 
Such properties are reflected in the components of an electric dipole
moment vector $\boldsymbol{\mu}$ that we associate to each mode
by multiplying the Cartesian displacement of
each atom by its effective 
charge and summing all the vectors thus obtained (see Table 3). This
gives an easy criterion for 
the correct identification of frequencies in the minimization
procedure, 
 i.e. in the search for the
parameters of the potential that minimize the function
\begin{equation}\label{scarto}
W=\frac{1}{16}\sum_i^{16}\left(\frac{\omega_i -
  \omega^{ex}_i}{\omega^{ex}_i}\right)^2\ ,
\end{equation}
where $\omega^{ex}_i$ are the experimental values.  This optimization
procedure was performed using a simulated annealing 
algorythm \cite{annealing},
especially useful for multivariate functions.  Actually this algorithm
provides many different solutions to the optimization problem,
possibly due to the presence of many local minima in the function that
has to be minimized.

Then, for any single set of parameters obtained we  determine
 the corresponding equilibrium position and the corresponding set 
of normal mode
frequencies.  The set of parameters is accepted if the calculated
equilibrium position is sufficiently near to the experimental one of
$\alpha$ quartz, the frequencies are sufficiently near to the
experimental ones, and furthermore the $\alpha$ structure is stable up
to sufficiently high temperatures.  No set of parameters found gave an
agreement for the frequencies better than about 6--7\%.

Other attempts were as follows.  We started from the power $n=6$
entering (\ref{potenziale}), letting $n$ be a free parameter,
different for each pair, adapting the cutoff parameter to each
choice. No substantial improvement was obtained. Then we changed
completely the form of the potentials, using Lennard--Jones ones. But
this gave a drastic worsening of the results.

These facts show that the results depend in a very sensitive way on
the form of the potentials.  In order to bypass this problem we
decided to restrict our studies to the linear model, assigning as
parameters directly the elastic constants. In such a way one even
eliminates the constraint that the elastic constants should be defined
in terms of first and second derivatives of a given potential.
However, no progress was obtained.

As a last resort, we eliminated the neutrality constraint
(\ref{neutro}) on the effective charge, by assuming both charges to be
free parameters, but again without substantial improvement.

Actually, in the whole literature we were unable to find a paper in
which the calculated frequencies agree with the experimental ones, in
the mean, better than 3\%, which is the result obtained in the old
paper \cite{altri2}.  In such a paper a linear model is considered,
which takes into account also the polarization of oxygen ions as a
free parameter. However, the maximum error was larger than 7\%.

In the more recent paper \cite{scandolo}, a nonlinear model was
investigated by MD simulations.  Both the short range potentials and
the ions polarizability were obtained through \emph{ab initio}
computations, but again, in the very words of the authors,
\textit{``The calculated frequencies are systematically
  underestimated, but differences are below 7\%--8\%.''}.  The authors
point out that their results constitute an improvement with respect to
those of previous works, for example those of paper \cite{Tse}, in
which \textit{``The discrepancy in the lower energy bending vibrations
  is somewhat higher, usually around 10--20\%''}. The authors of
\cite{scandolo} ascribe the improvement to the fact of having taken
the ions' polarization into account.
   

Neither does the consideration of three--body potentials, apparently,
improve the agreement between computed and observed frequencies. For
example, in paper \cite{Ma} the errors can reach 20\% for some modes
(see their Table V), as also occurs in paper \cite{Kieffer} where for
example (see fig. 10) one of the computed lines lies near 600
cm$^{-1}$, against the observed value near to 500 cm$^{-1}$.

We interpret these facts as indicating that some structural deficiency
is present in all models (including ours) that have been
considered. Such a deficiency was sometimes acknowledged. For example,
in paper \cite{DellaValle} it is stated that
\textit{``The deviations of the
calculation with respect to the experiment are not random,
but systematic: the higher phonon frequencies, above the
mean, are invariably too low with respect to the experiment,
while the lower frequencies are invariably too high.''}. These facts
are  usually ascribed to some deficiency in the short range potentials.
We conjecture instead that it is the way of dealing with the
long--range forces that should play a particularly relevant role in
this connection. Indeed, the experimental measures suggest the
existence of a splitting between the longitudinal modes and the
transverse ones, that should be due to the long--range forces.
On the other hand this splitting is actually the feature that the
considered models fail to properly describe.  We intend to come back
to this point in a further work.


\acknowledgments We thank G. Grosso, G. Pastori Parravicini, N. Manini
and G. Onida for useful discussions. The use of computing resources
provided by CINECA is also gratefully acknowledged.

\begin{table}[h]
  \begin{center}
    \begin{tabular}{cccccc}
     \hline 
      sym. & exp. & calc. & $\mu_x$ & $\mu_y$ & $\mu_z$\\
      & freq. & freq. & & & \\
      \hline 
      &\ 1162\ \ & \ 1125\ \ & \ -0.1947\ \ & \ 0.1903\ \ & 0\\
      & & 1125 & -0.1903 & -0.1947 & 0\\
      &1072 & 1083 & 0.3479 & -0.4105 & 0\\
      & & 1083 & -0.4105 & -0.3479 & 0\\
      &795 & 727 & 0.3032 & -0.1175 & 0\\
      && 727 & -0.1175 & -0.3032 & 0\\
      &697 & 675 & 0.1058 & 0.1061 & 0\\
      & & 675 & -0.1061 & 0.1058 & 0\\
      $E$ & 450 & 506 & 0.4470 & 0.2172 & 0\\
      & & 506 & 0.2172 & -0.4470 & 0\\
      & 394 & 359 & -0.3075 & 0.0309 & 0\\
      & & 359 & -0.0309 & -0.3075 & 0\\
      & 265 & 258 & -0.0010 & 0.0010 & 0\\
      & & 258 & 0.0010 & 0.0010 & 0\\
      & 128 & 141 & 0.0435 & 0.0072 & 0\\
      & & 141 & 0.0072 & -0.0435 & 0\\
      \hline
      & 1080 & 1101 & 0 & 0 & -0.5871\\
      $A_2$ & 778 & 732 & 0 & 0 & 0.3754\\
      & 495 & 544 & 0 & 0 & 0.4078\\
      & 364 & 358 & 0 & 0 & -0.4294\\
      \hline
      & 1085 & 1074 & 0 & 0 & 0\\
      $A_1$ & 464 & 474 & 0 & 0 & 0\\
      & 356 & 355 & 0 & 0 & 0\\
      & 207 & 232 & 0 & 0 & 0\\
      \hline 
    \end{tabular}
  \end{center}
  \caption{Normal modes of $\alpha$ quartz grouped by
    symmetry. Experimental frequencies (from \cite{scottporto}) and calculated frequencies 
    in cm$^{-1}$ are reported, together with the calculated components of
    the dipole moment $\boldsymbol{\mu}$ for each mode. }
  \label{tab:3}
\end{table}

\end{document}